\documentclass[twocolumn, aps, 10pt, pra, longbibliography,floatfix]{revtex4-2}

\usepackage{soul}
\usepackage{amsmath}   
\usepackage{amssymb}
\usepackage{mathtools}
\usepackage{graphicx}
\usepackage{dcolumn}
\usepackage{bm}
\usepackage{amsmath}
\usepackage{graphicx} 
\usepackage{amsfonts}
\usepackage{subfigure}
\usepackage{graphicx}
\usepackage{array}
\usepackage{color}
\usepackage{amssymb}%
\usepackage{makecell}
\usepackage[colorlinks=true,linkcolor=blue]{hyperref}%
\hypersetup{allcolors=blue}
\usepackage[normalem]{ulem}
\usepackage{xcolor}
\usepackage{tabularx,booktabs}
\usepackage{multirow}
\usepackage{threeparttable}

\usepackage{mathtools}
\DeclarePairedDelimiterX\braket[2]{\langle}{\rangle}{#1 \delimsize\vert #2}

\newcommand{\micro}[1]{\ensuremath{\mu}{#1}}

\newcolumntype{Y}{>{\centering\arraybackslash}X}

\begin{document}

\title{Electric field diagnostics in a continuous rf plasma using Rydberg-EIT}
\author{Bineet Dash}
\email{bkdash@umich.edu}
\affiliation{Department of Physics, University of Michigan, Ann Arbor, MI 48109, USA} 

\author{Xinyan Xiang}
\affiliation{Department of Physics, University of Michigan, Ann Arbor, MI 48109, USA} 
\author{Dingkun Feng}
\affiliation{Department of Physics, University of Michigan, Ann Arbor, MI 48109, USA} 
\author{Eric Paradis}
\affiliation{Department of Physics \& Astronomy, Eastern Michigan University, Ypsilanti, MI 48197, USA}
\author{Georg Raithel}
\affiliation{Department of Physics, University of Michigan, Ann Arbor, MI 48109, USA}

\begin{abstract}
We present a non-invasive spectroscopic technique to measure electric fields in plasma, leveraging large polarizabilities and Stark shifts of Rydberg atoms. Rydberg Stark shifts are measured with high precision using narrow-linewidth lasers via Electromagnetically Induced Transparency (EIT) of rubidium vapor seeded into a
continuous, inductively coupled radio-frequency (rf) plasma in a few mTorr of argon gas. Without plasma, the Rydberg-EIT spectra exhibit rf modulation sidebands caused by electric- and magnetic-dipole transitions in the rf drive coil. With the plasma present, the rf modulation sidebands vanish due to screening of the rf drive field from the plasma interior. The lineshapes of the EIT spectra in the plasma reflect the plasma's Holtsmark microfield distribution, allowing us to determine plasma density and collisional line broadening over a range of pressures and rf drive powers. The work is expected to have applications in non-invasive spatio-temporal electric-field diagnostics of low-pressure plasma, plasma sheaths, process plasma and dusty plasma.

\end{abstract}

\date{\today }
\maketitle

\section{Introduction}
\label{sec:intro} 

Knowledge of the spatial and temporal structure of electric fields within plasma, particularly near sheaths and boundaries, is paramount for accurate modeling of transport processes, energy transfer, and particle confinement mechanisms. Conventional plasma diagnostic methods employing electrostatic metal probes are fundamentally limited by back action of the probe on the local plasma state, and often lack the necessary spatial resolution. State-of-the-art non-perturbative spectroscopic techniques~\cite{Goldberg_Hoder_Brandenburg_2022}, such as Electric Field Induced Second Harmonic generation (E-FISH) ~\cite{Dogariu_2017_EFISH_1, Goldberg_2018_EFISH_3, simeni2018_EFISH_4} and Coherent Anti-Stokes Raman Scattering (E-CARS)~\cite{Lempert_ecars2, Goldberg_2015_ecars2}, exploit plasma-driven nonlinear susceptibilities, while passive methods like Optical Emission Spectroscopy (OES)~\cite{brose_oes1, griem2005principles_oes} infer fields from the Stark broadening of emission line profiles. These techniques typically require strong electric fields, often on the order of several $\text{~kV/cm}$, thereby restricting their practical application to intermediate or near-atmospheric-pressure discharges.

In contrast, low-temperature rf discharges generated at pressures below $100 \text{~mTorr}$ are characterized by weak internal electric fields, typically in the few $\text{V/cm}$ range. These low-pressure plasmas are critical for a wide array of applied technologies including semiconductor processing, biomedical applications, and advanced propulsion systems~\cite{chu2013low, mazouffre2012laser}, as well as for fundamental studies of magnetized plasmas~\cite{popov2016advances_magnetizedplasma} and dusty plasmas~\cite{melzer2021_magnetizeddustyplasma}. Spatial mapping of these weak fields with existing probe-based or optical methods remains a challenge due to insufficient field sensitivity. Electric fields above $40 \text{~V/cm}$ in low-pressure discharges have been measured using laser-induced florescence of field-mixed molecular levels~\cite{Moore_Davis_Gottscho_1984} and optogalvanic laser spectroscopy~\cite{shoemaker_1988}.

Rydberg atoms, {\sl{i.e.},} atoms with large principal quantum number $n$, are well-suited for electric field measurement via the Stark effect due to their exaggerated dipole polarizabilities that scale as $n^7$. The Stark effect of Rydberg atoms has been used to characterize electric fields in cold Rb$^+$ ion clouds~\cite{Feldbaum2002, Duspayev2023_highell} and in photoexcited plasma~\cite{weller19_interplay, dave2017_plasmaeit}. The Laser-Induced Fluorescence (LIF)-dip technique~\cite{Czarnetzki_1999_lifdip, Czarnetzki_1999_lifdip, Takizawa_2004_lifdip, Takizawa_2007_lifdip}, which monitors the loss of fluorescence from one of the involved states during two-photon excitation, has extended sensitivity.  However, its performance is limited by the reliance on semiclassical population transfer to obtain the dip and by the technical difficulty of detecting a small signal reduction against a large baseline fluorescence.

Electromagnetically Induced Transparency (EIT)~\cite{sedlacek12}, a coherent quantum optical effect, enables high-resolution narrow-linewidth spectroscopy of Rydberg levels. EIT arises in two-photon excitation of Rydberg states via a resonant, decaying intermediate state, wherein interference of excitation pathways into that state creates a narrow spectral window of transparency for the probe laser~\cite{finkelstein2023practicaleit}. Rydberg-EIT allows the measurement of Rydberg-level Stark shifts using the absorption of EIT probe light as readout. Rydberg-EIT in alkali vapor cells has emerged as a robust quantum-sensing methodology for dc~\cite{ma2020dc} and radio frequency (rf) electrometry~\cite{holloway14, receiver2021}. The viability of Rydberg-EIT was demonstrated in inert background gases of up to several $\text{Torr}$~\cite{thaicharoen2024}. Below $\sim 100 \text{~mTorr}$ of inert-gas pressure, shifts and distortions of Rydberg-EIT lines remain quite minimal~\cite{dash2025rydberg}. These results opened up Rydberg-EIT as a promising venue for field diagnostics in low-pressure discharge plasmas in inert gases.

In the present paper, we apply Rydberg-EIT to analyze electric fields in inductively coupled rf plasmas (ICPs) in a few $\text{mTorr}$ of argon. Rydberg-EIT spectroscopy is performed on a trace amount of rubidium vapor immersed in the ICPs. Our results provide direct evidence of effective screening of the plasma rf drive field from the interior of the plasma volume. Further, the method allows a spectroscopic measurement of the plasma microfield distribution in cw rf plasma with microfields ranging from {$\sim 5 \text{~V/cm}$} to $\sim 10 \text{~V/cm}$. Our results validate the feasibility of EIT in a dynamic rf plasma environment and underscore the potential of Rydberg-EIT in spatio-temporal electric-field mapping in bulk plasma, in plasma sheaths, and near solid impurities, including dust.

\section{Experimental Setup}
\label{sec:exp}

\begin{figure*}
    \centering
    \includegraphics[width=0.85\linewidth]{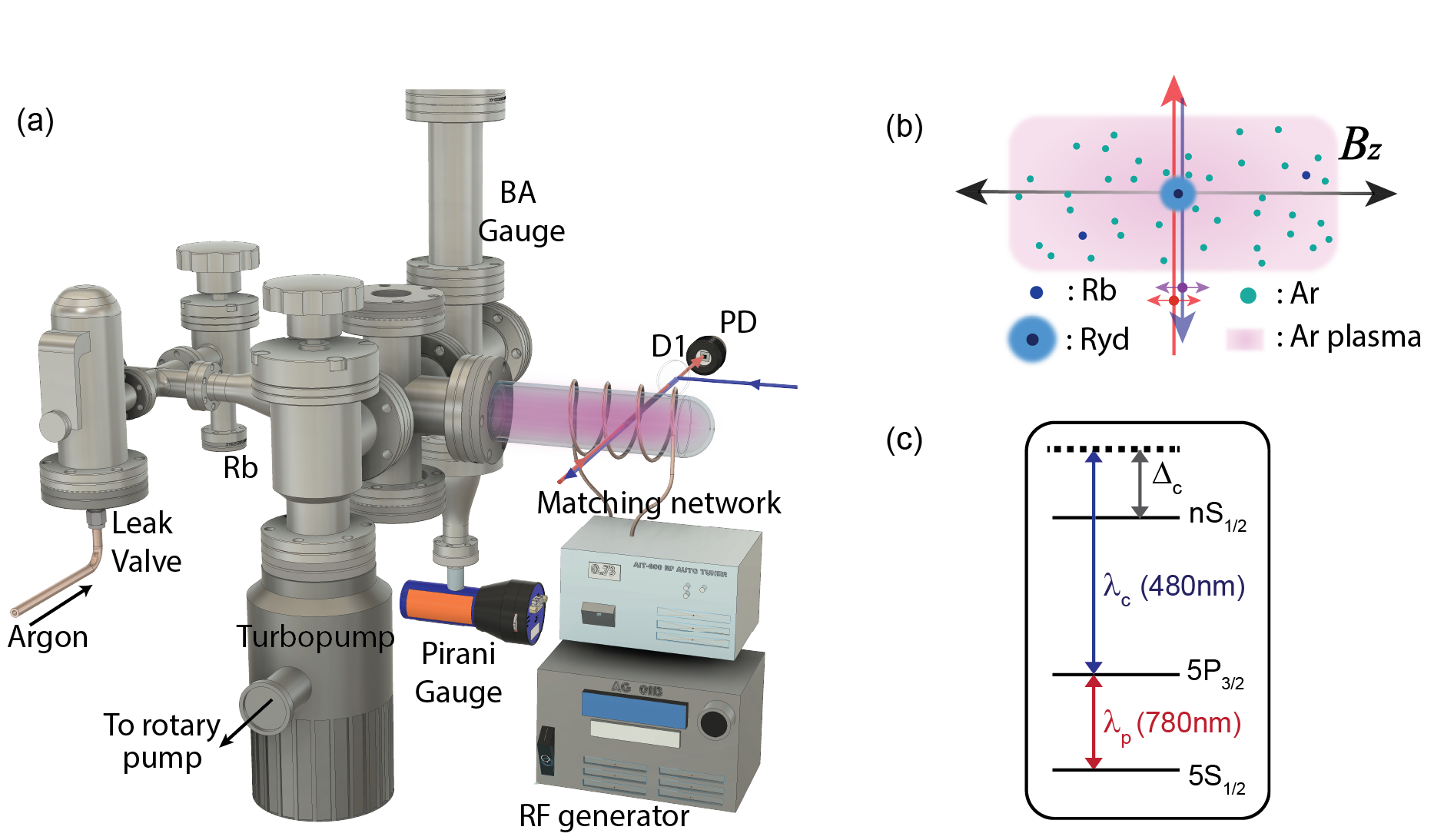}
    \caption{(a) Schematic of the experimental setup for plasma discharge and EIT spectroscopy. An inductively coupled Ar plasma (pressure range $\sim 5$ to 50~mTorr) is generated inside the depicted domed glass cylinder by a 13.56~MHz rf source, a matching network, and a helical drive coil. Near-thermal Rb vapor, entrained into the Ar flow and the plasma discharge, is optically interrogated for Rydberg-EIT-based electric-field diagnostics. (b) Illustration of the discharge volume and EIT interrogation region. The 780-nm EIT probe (red) and 480-nm coupling laser beams (purple), combined with a dichroic mirror (D1), counter-propagate and overlap each other inside the domed glass cylinder. The transmitted EIT probe power is measured with a photodiode (PD). Both laser beams are linearly polarized parallel to the rf drive magnetic field, $B_z$. (c) Rb energy levels used in this work. The probe laser (wavelength $\lambda_p = 780$~nm) is on-resonance with the $5S_{1/2}, F=3$ $\leftrightarrow$ $5P_{3/2}, F'=4$  transition, while the coupling laser (wavelength $\lambda_c = 480$~nm) is scanned across a $5P_{3/2}, F'=4 \leftrightarrow nS_{1/2}$ transition for Rydberg-EIT-based plasma electric field diagnostics.}
    \label{fig:setup}
\end{figure*}

The plasma discharge and EIT spectroscopy take place in the ultrahigh vacuum (UHV) chamber shown in Fig.~\ref{fig:setup} (a). This chamber contains a mixture of pure Ar gas, which serves as the plasma precursor, and a trace of Rb vapor near its thermal pressure at room temperature. The Rb vapor is optically interrogated for Rydberg-EIT-based electric field sensing. Sustained plasma operation requires a continuous and controlled flow of pure Ar gas through the chamber, while maintaining the desired Ar pressure. Pure Ar from a commercial gas cylinder is supplied into a service line at 1.5 $\text{atm}$. The overpressure prevents back-diffusion of atmospheric contaminants into the service line. The Ar flows via an Alicat mass flow controller (MFC-20 sccm) held at positive pressure, and then through a UHV leak valve into the plasma chamber. The MFC provides precise Ar flow rate control, while the leak valve provides a UHV-compatible vacuum break. Downstream of the leak valve, the Ar flow is merged with a Rb vapor stream from a Rb reservoir, which is pre-filled with $\sim 1$~g of natural isotopic Rb.

The UHV chamber is evacuated by a $70 \text{~L/s}$ turbomolecular pump (Varian V-70), which connects via a right-angle valve to a $1.6 \text{~L/s}$ rotary pump. The plasma chamber is initially evacuated to a base pressure of $5 \times 10^{-7} \text{~Torr}$, as measured by a Bayard-Alpert (BA) ionization gauge. When Ar is flowed through the system, the pressure is measured with a Pirani gauge. We study ICPs with Ar pressures ranging from $\sim 5 \text{~mTorr}$ to $50 \text{~mTorr}$. The Ar pressure is regulated by balancing the flow through the leak valve, which is metered by the MFC, against the effective pumping speed, which is throttled down by partially closing the right-angle valve. The throttling keeps the Ar flow within a range of 0.5~sccm to 10~sccm, which is sufficiently low to maintain an equilibrium Rb pressure near $3 \times 10^{-7} \text{~Torr}$ within the Ar, approximately equivalent to the thermal Rb vapor pressure at room temperature ($\approx 295 \text{~K}$). The Rb pressure is monitored via the absorption coefficient of the $^{85}$Rb vapor on the 780-nm D$_2$ line observed in the domed glass cylinder in Fig.~\ref{fig:setup}~(a). After heating of the Rb reservoir to 45 $^\circ$C, the injected Rb-vapor flow suffices to compensate for the continuous removal of Rb vapor entrained within the Ar flow system. In this way, the Rb atom number density in the Rydberg-EIT field sensing region in the domed glass cylinder in Fig.~\ref{fig:setup}~(a) is maintained near its thermal value at room-temperature.

The ICP discharge is generated within a domed glass cylinder measuring $4~\text{cm}$ in diameter and $10~\text{cm}$ in length appended to the UHV chamber. The rf power, $P_{\text{rf}}$, at 13.56~MHz is supplied by a commercial generator (T\&C Power Conversion, model AG-0113) and delivered via a matching network to a 3-turn helical rf drive coil of 5~cm diameter, constructed from AWG 11 copper wire. The rf magnetic field produced by this coil couples through the domed glass cylinder into the ICP. The rf magnetic field under absence of plasma extends along the axis of the glass cylinder and amounts to about 0.8~G/A. 

Impedance matching is accomplished by a capacitive matching network (T\&C Power Conversion, model AIT-600), which ensures efficient power transfer into the inductive load and keeps the reflected power below 5\%. Photoelectrons are seeded into the Ar gas by illuminating the glass cylinder with UV light before ICP ignition. The seeding mechanism facilitates stable ICP operation at 
{$P_{\text{rf}}$
as low} as 0.15~W. We keep the rf power $\lesssim 5$~W to avoid Rb pressure loss. At higher rf powers, and with the ICP present, we have observed a rapid loss of Rb pressure from the Rydberg-EIT region. We attribute the Rb loss to a combination of plasma etching on the inside glass surface and subsequent surface chemistry that eliminates metallic Rb adsorbed on the glass. Loss of a thin layer of Rb adsorbate on the glass is believed to be the cause for Rb pressure loss at higher rf powers.

EIT-based plasma diagnostics are performed with cascade EIT from the Rb ground state, $5S_{1/2}$, to an $nS_{1/2}$ Rydberg state through the intermediate $5P_{3/2}$ state, as depicted in Fig.~\ref{fig:setup}~(c). This is achieved by counter propagating 780~nm and 480~nm EIT probe and coupler laser beams through the discharge volume, as shown in Fig.~\ref{fig:setup}~(b). The laser beams have respective powers of $\approx 3.4$~\micro{W} and 35~mW, are focused at the center of the ICP volume in the domed glass cylinder with Gaussian beam-waist parameters  of $w_0=60~\mu$m and $75~\mu$m, and are
linearly polarized with polarizations parallel to the axis of the helical rf drive coil.
The Rabi frequencies for the EIT probe (lower) transition range between $\sim 12$~MHz and $\sim 20$~MHz, and those for the EIT coupler (upper) transition into $25S_{1/2}$ between
$\sim 8$~MHz and $\sim 14$~MHz, for the magnetic quantum number $m_F$ ranging from 3 to 0.
The frequency of the 780~nm EIT probe laser is locked to the lower $F=3$ to $F'=4$ hyperfine transition. The transmitted EIT probe power is acquired using a photodiode (FDS-100), which is protected from plasma fluorescence and other stray light by a 780~nm band-pass filter. The EIT probe power is recorded as a function of the frequency of the 480~nm EIT coupler beam, which is scanned across the Rydberg transition. The coupler-laser frequency scans are calibrated and linearized using the transmission spectrum of a low-power beam sample passed through a Fabry-Perot reference cavity, which has a stability on the order of 10~MHz per 24 hours~\cite{hansis2005}. In our data, zero coupler detuning marks the field-free Rydberg-EIT resonance between the intermediate $5P_{3/2}, F'=3$ level and the target Rydberg level with the Ar gas present, which causes small shifts~\cite{dash2025rydberg}. The Rydberg-EIT spectra, {\sl{i.e.}}, the transmitted EIT probe power measured as a function of coupler detuning, $\Delta_{c}$, then reflect the accumulated effects of Zeeman and Stark shifts averaged over the length of the EIT probe's path through the glass cylinder in Fig.~\ref{fig:setup}.   

The number density of the Rb tracer atoms in the ICP is about five orders of magnitude lower than that of Ar. Hence, the Rb electric-field-sensing atoms are presumed to have no significant effect on the characteristics of the Ar ICP under study, beyond an observed attenuation effect of Rb on the glass walls on the RF coupling efficiency into the ICP.

\begin{figure*}
    \centering
    \includegraphics[width=0.95\linewidth]{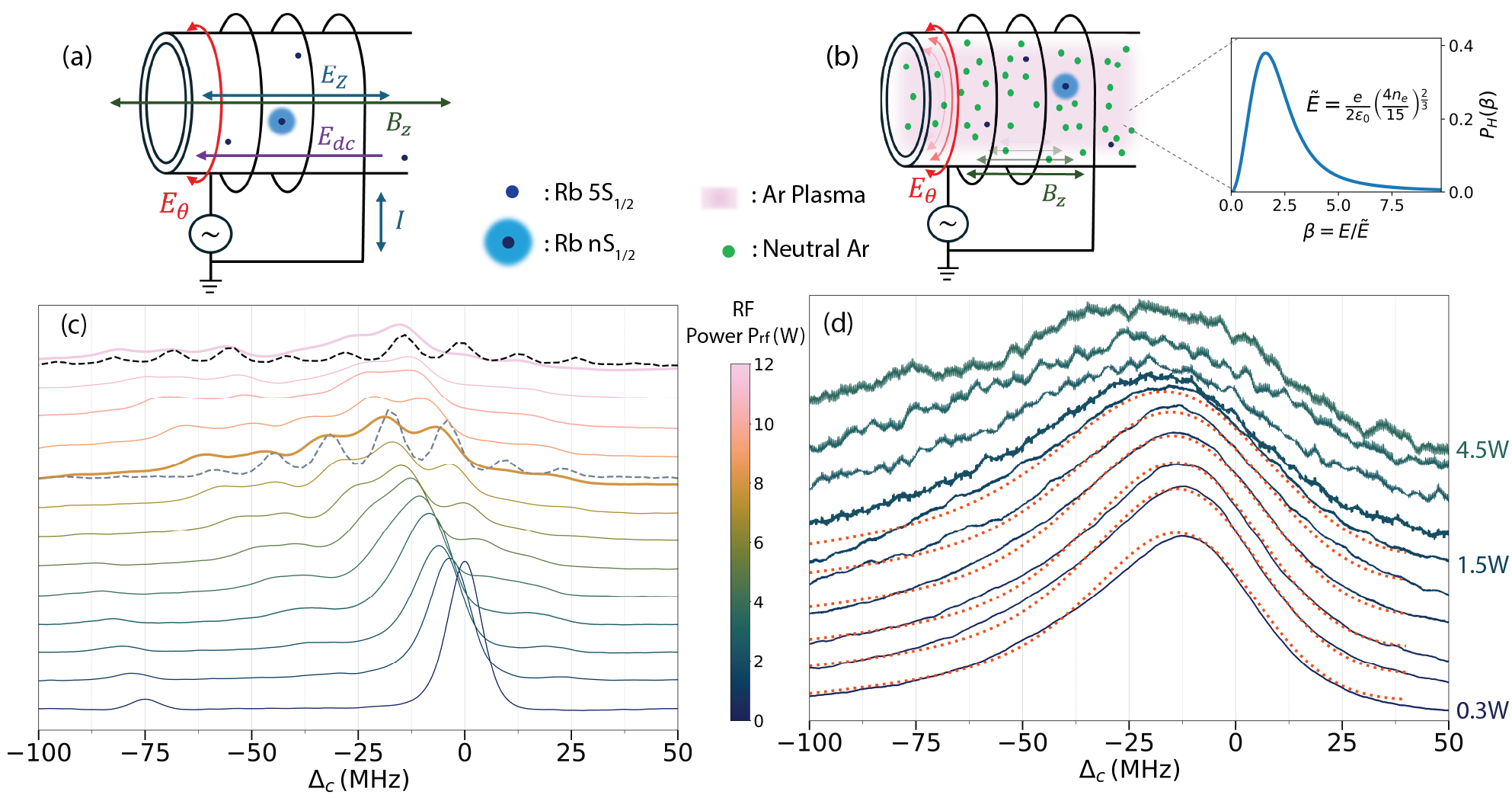}
    \caption{(a) Fields arising from applied rf power, $P_{\text{rf}}$, at 13.56~MHz in the chamber in the absence of Ar: axial rf magnetic field $B_z$, induced rf electric field $E_{\theta}$, capacitive rf electric field $E_z$, and dc electric field $E_{dc}$ (self-bias). (b) Within Ar plasma, the rf fields are shielded due to the skin effect, leaving the stochastic microfield as the dominant field. The microfield distribution is given by the Holtsmark distribution (inset). (c) Measured Rydberg-EIT spectra (solid lines) of Rb $27S_{1/2}$ without plasma, with $P_{\text{rf}}$ indicated by the color bar. The rf drive fields create electric and magnetic rf sidebands at integer multiples of $\omega_{\text{rf}}$. 
    %The sidebands are shifted and are modulated in strength following polarization-dependent patterns.
    Calculated spectra, shown in gray (black) dashed lines, are for EIT-beam polarizations along $z$, $B_z$ = 7.5~G (15~G), $E_{dc}$=2~V/cm (4~V/cm), and $E = \sqrt{E_\theta^2 + E_z^2}= 10~$V/cm (12~V/cm). Aside from line broadening likely caused by field inhomogeneities along the EIT probe beam, the simulated spectra are a fairly close match to experimental spectra at $P_{\text{rf}}= 8$~W (12~W). (d) Rydberg-EIT spectra in plasma with 12~mTorr of Ar. The rf-induced sidebands disappear due to the skin effect. The asymmetric line shape is due to the Rydberg-atom Stark effect in the plasma's microfields, which follow a Holtsmark distribution. The red dashed curves show our fit results obtained from fits according to Eq.~\ref{eq:all_conv}. Identical color scaling is used in panel (c) and (d). 
    }
    \label{fig:fig2}
\end{figure*}

\section{Rydberg-EIT spectra in ICP antenna without plasma}
\label{sec:plasma-free}

We first investigate Rydberg-EIT spectra of Rb vapor within the fields generated by the rf drive coil under Ar-free conditions, for a set of rf powers. The drive frequency of 13.56~MHz, which is in the High Frequency (HF) radio band, generates electric and magnetic Rydberg level shifts that are quasi-static on intrinsic atomic time scale. These shifts manifest in the form of frequency modulation sidebands as well as AC Stark shifts in Rydberg-EIT spectra~\cite{Miller2016, Bason2010}. In the absence of dc electric fields and rf magnetic fields, only even harmonics of the drive field's frequency occur. The sideband structures and shifts observed inside solenoidal rf coils, shown in Fig.~\ref{fig:fig2}~(c), are considerably richer and include both even- and odd-harmonic rf bands. 

The complexity of Rydberg-EIT spectra in solenoidal rf coils is due to the presence of several types of fields interacting with the atoms, which are indicated in Fig. \ref{fig:fig2}~(a). The rf current through the helical coil generates an axial rf magnetic field with amplitude $B_z$, which induces linear rf Zeeman shifts in all atomic levels involved. Further, the induced azimuthal rf electric field with amplitude $E_\theta$ induces quadratic rf Stark shifts in the utilized Rydberg levels. Also, the reactive voltage drop across the length of the drive coil, which can reach hundreds of V, couples capacitively through the glass walls and gives rise to an axial rf electric field with amplitude $E_z$, which appears in-phase with the azimuthal induced field $E_\theta$. Notably, $E_z$ is approximately homogeneous, while $E_\theta$ scales with the radial separation from the coil axis. Finally, in our asymmetric driving configuration, in which one coil terminal is grounded and the other is driven by the rf potential, a longitudinal dc self-bias field~\cite{Sugai_1994_selfbias}, $E_{dc}$, develops due to vastly different electron and ion masses. Generally, highly mobile negative charges are removed near the polarized terminal of the drive coil, resulting in a longitudinal charge distribution within the cell, which in turn causes the dc self-bias electric field, $E_{dc}$.

%moved some text around based on feedback
Rydberg-EIT in spectroscopic cells located inside rf coils differs greatly from Rydberg-EIT in freely propagating rf fields, where magnetic-field effects are typically negligible~\cite{Miller2016, Bason2010}. The magnetic field within an rf drive coil is several orders of magnitude higher than that in traveling electromagnetic waves in source-free regions, leading to effects from rf Zeeman shifts that rival those of Rydberg Stark shifts, and therefore must be included in the analysis. 
%In rf coils, the effects of rf Zeeman shifts can rival those of rf and dc Rydberg Stark shifts.   
The magnetic-dipole Zeeman couplings are linear in $B_z$ and generate rf sidebands on all atomic levels with non-zero magnetic moments. Zeeman sidebands have shifts of $k\omega_{\text{rf}}$ from the unperturbed atomic levels and relative amplitudes of ${\rm{i}}^k J_k (\frac{\mu_B g_F B_z m_F }{\hbar \omega_{\text{rf}}})$ with integers $k$, Lande factors $g_F$ and magnetic quantum numbers $m_F$.
% The Lande factors $g_F$ depend on $n$, $\ell$, $J$, and $F$, the principal, orbital, electronic, and hyperfine quantum numbers of the $5S_{1/2}$, $5P_{3/2}$, and $nS_{1/2}$ levels.

Second-order electric-dipole couplings give rise to the quadratic Stark effect of the electric-field-sensitive Rydberg levels. The Stark rf sidebands have shifts of $-\frac{\alpha E^2}{4} + 2 k \hbar \omega_{\text{rf}}$ and amplitudes $(-1)^k J_k ( \frac{\alpha E^2}{8 \hbar \omega_{\text{rf}}})$, where $\alpha$ is the $nS_{1/2}$ Rydberg state's scalar electric polarizability and $E = \sqrt{E_z^2 + E_\theta^2}$, the total rf electric field~\cite{Miller2016, Bason2010}. Electric rf fields generate even sidebands, while magnetic rf fields generate both even and odd rf sidebands, with amplitudes of coinciding bands adding coherently. The longitudinal dc self-bias field, $E_{dc}$, adds an additional dc Stark shift of $-\frac{\alpha E_{dc}^2}{2}$ to all Rydberg rf sidebands, and it coherently adds mixed ac-dc terms to their amplitudes. The complicated modulation structure in the rf drive coil follows conveniently and naturally from our direct, time-dependent model explained next.

The EIT spectra shown in Fig.~\ref{fig:fig2}~(c) are reproduced in their essential parts by a simulation of the time-dependent Lindblad equation. The optical EIT fields in the simulation are linearly polarized along the rf coil axis. Since the induced electric and the magnetic rf drive fields, $E_\theta$ and $B_z$, are orthogonally polarized, all $m_F$-substates with the same $n$, $\ell$, $J$, and $F$ quantum numbers are generally coupled. For $^{85}$Rb, there are 48 coupled atomic levels. We include all Zeeman and Stark couplings arising from the aforementioned fields, as well as detailed rates for spontaneous decays, in a time-dependent simulation. All drive matrix elements and decay rates carry the proper dependencies on all involved angular quantum numbers.

Due to the large number of states, the time-dependent Lindblad equation is solved using the quantum trajectory method, which yields an approximation to the density matrix and the atomic absorption coefficient by averaging over a large number of stochastic wave-functions~\cite{Dalibard1992, Molmer1993, Zhang2018, Xue2019}. The quantum trajectories are typically propagated over $2~$\micro{s} of evolution time for each trajectory, which is on the order of the time of flight of a typical atom through the EIT probe region. The Hamiltonian includes constant level shifts, field detunings and Doppler shifts, as well as the aforementioned time-dependent magnetic and electric rf level shifts, which oscillate at frequencies of $\omega_{\text{rf}}$ or $2 \omega_{\text{rf}}$. 
%The rf Stark shift of the $nS_{1/2}$ Rydberg states due to the in-phase, orthogonally polarized rf electric fields is $ - (\frac{\alpha}{4}) (E_z^2 + E_\theta^2) (1+ \cos (2 \omega_{\text{rf}} t))$. 
Further, the EIT probe absorption is integrated over the Maxwell velocity distribution of the Rb vapor. 

The simulation examples in Fig.~\ref{fig:fig2}~(c), denoted by gray and black dashed lines for the cases  $P_{\mathrm{rf}}$ = 8 W and 12 W respectively, show good qualitative agreement with measured plasma-free rf spectra in terms of locations and relative strengths of rf sidebands. Deviations are attributed to field inhomogeneities. In view of the results discussed next, the plasma-free EIT spectra in the rf drive coil in Fig.~\ref{fig:fig2}~(c) serve as baselines that allow us to determine to what extent the plasma surface shields the rf field from the plasma's interior volume, where most of the EIT signal originates from.

\section{Rydberg-EIT spectra in plasma}
\label{sec:plasma-EIT}

Rydberg-EIT spectra of Rb tracer vapor immersed in Ar ICPs were recorded over a range of applied rf powers, $P_{\text{rf}}$, of up to about 5~W and Ar pressures ranging from $\sim 5$ to $50 \text{~mTorr}$. Selected EIT spectra in plasmas generated from $12 \text{~mTorr}$ of Ar are shown in Fig.~\ref{fig:fig2}~(d). The presented Rydberg-EIT spectra in the ICP plasma show a striking disappearance of the rf sidebands seen in Fig.~\ref{fig:fig2}~(c). The absence of rf sidebands demonstrates the efficient screening of electric and magnetic rf fields from the interior of the plasma volume, which is attributed to the skin effect. Further, even with a continuous rf ICP present, high-contrast Rydberg-EIT lines of the Rb tracer vapor remain, indicating considerable resilience of the Rb vapor against depletion that may result from plasma-induced ionization {of Rb}. The EIT lines in Fig.~\ref{fig:fig2}~(d) exhibit an asymmetric lineshape, indicative of inhomogeneous electric-field broadening inside the plasma. As $P_{\text{rf}}$ is increased, the EIT lines experience larger broadening, along with an increasing red shift. Both features indicate an increase in electron density and plasma microfield strength. In the following, the EIT lineshapes shown in Fig.~\ref{fig:fig2}~(d) are analyzed considering inhomogeneous Stark shifts and collisional broadening of the Rydberg-EIT line in the plasma environment. 

\begin{figure}[b!]
    \centering
    \includegraphics[width=0.95\linewidth]{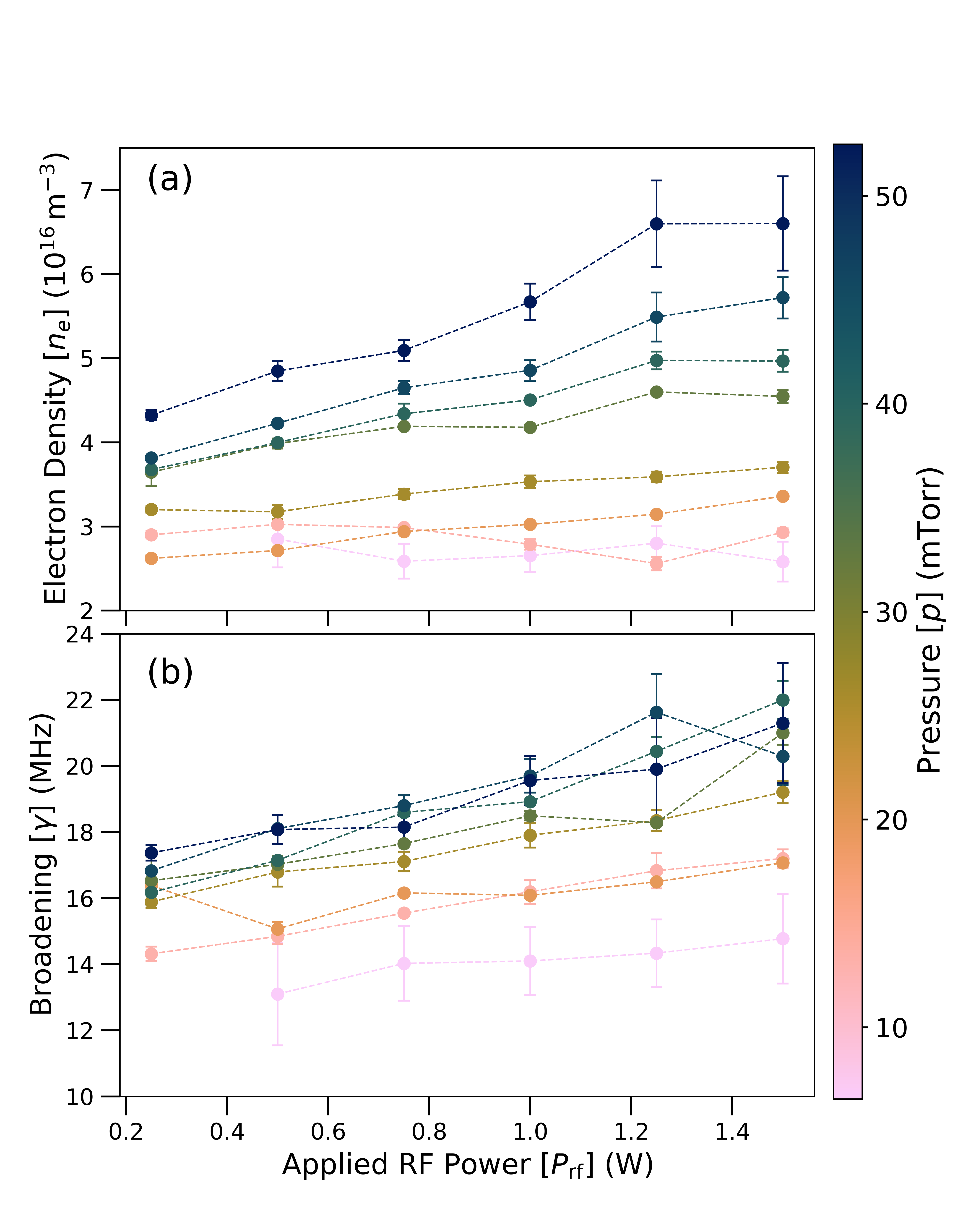}
    \caption{ Plasma parameters extracted from Rydberg-EIT spectra for $25S_{1/2}$, obtained by fitting based on the the three-way convolution in Eq.~(\ref{eq:all_conv}), plotted vs $P_{\rm{rf}}$
    with color-coded Ar pressure $p$. (a) Plasma density $n_e$ extracted from the Holtsmark part of the fits, assuming quasi-neutrality in the bulk ($n_e \approx n_i$).  Over the $n_e$-range found, the most probable microfield ranges from 5.1~V/cm to 9.8~V/cm. At $p \gtrsim 25$~mTorr, the slopes $\Delta  n_e / \Delta P_{\mathrm{rf}}$ undergo a significant increase.
    (b) Broadening $\gamma$ from collision-induced dephasing mechanisms. Error bars exceed statistical fit uncertainties and rather reflect confidence based on signal to noise ratio and drift in the underlying EIT spectra.}
    \label{fig:fig3}
\end{figure}

The interpretation of the Rydberg-EIT spectra in our ICP system {is simplified} by several factors. The dominant portion of the EIT signal results from the overlap of the EIT probe and coupler beam within the bulk discharge volume, where macroscopic quasineutrality is maintained ({\sl{i.e.}}, electron density $n_e \approx n_i$, ion density). Furthermore, the plasma in question is a low-temperature plasma (LTP), with characteristic electron energies  and ion temperatures ranging from 1~eV to 3~eV \cite{Lee_Chung_2011, donnelly2004plasma_Te_review} and 500~K to 800~K~\cite{Hebner_1996_Tion}, respectively. In our case, EIT has a dynamic time scale $\tau_{\text{EIT}}$ on the order of a few tens of ns~\cite{Scully1997}, corresponding to only a few $\mu$m of ion travel. Hence, the ions are near-stationary during $\tau_{\text{EIT}}$ and generate a stochastic electric-field distribution at the location of the Rydberg atoms, known as the plasma microfield (Holtsmark) distribution~\cite{Holtsmark1919,griem2005principles_oes}. Conversely, the much faster electrons primarily give rise to collisional Rydberg-level dephasing, while their contribution to the electric microfield, averaged over $\tau_{\text{EIT}}$, is negligible. The lower levels $5S_{1/2}$ and $5P_{3/2}$ are not significantly shifted or broadened by the plasma, as determined by saturation spectroscopy of the Rb $D_2$ line in the plasma. Therefore, shifts and broadening of the EIT lines are solely attributed to inhomogeneous Rydberg Stark shifts and homogeneous collisional broadening of the Rydberg levels in the plasma.

%the Rydberg-atom-plasma interaction. 
%These can be categorized into two main components: inhomogeneous Stark shifts and inhomogeneous broadening from the ionic microfields, and a homogeneous collisional broadening from Ar$^+$ and e$^-$ in the plasma.

%The treatment of microfield-induced inhomogeneous broadening is simplified due to two factors. Firstly, 
% The $nS_{1/2}$ field-sensing Rydberg states have only a single allowed value of $|m_j|$, namely $|m_j| = 1/2$. Thus, the quadratic Stark shifts are solely given by a scalar polarizability, $\alpha$, allowing a simple mapping of the Rydberg Stark shift $\delta$ onto electric field via $\delta = -\frac{\alpha E^2}{2}$. 
Further, the plasma is only weakly ionized, ensuring that correlation effects between charged particles are negligible. Hence, the stochastic microfield generated by the ions is described by the Holtsmark distribution~\cite{Holtsmark1919, Demura_2010}  $\mathcal{P}_H(E)$, which depends on the plasma density, $n_e=n_i$:
\begin{equation}
    \mathcal{P}_{H}(E, n_e) = \frac{2}{\pi} \frac{E}{\tilde{E}^2} \int_0^\infty dx \, x \sin\left(\frac{E}{\tilde{E}} x\right) \exp\left(-x^{3/2}\right)
    \label{eq:holtsmark-pdf}
\end{equation}
\noindent where $\tilde{E} = \frac{e}{2\epsilon_0}\left(\frac{4n_e}{15}\right)^{\frac{2}{3}}$.  The most probable microfield is given by $1.61\tilde{E}$ and scales with plasma density as $(n_e)^{{2}/{3}}$. 

An $nS_{1/2}$ Rydberg level, possessing a quasi-static, scalar polarizability $\alpha > 0$, experiences a frequency shift $\delta = -\frac{1}{2} \alpha E^2$ due to the quadratic Stark effect. Consequently, the plasma microfield effect manifests as a convolution of the field-free EIT lineshape $I_0(\Delta_c)$ with the probability distribution of the Holtsmark-induced line shift, $\tilde{\mathcal{P}}_H (\delta, n_e) = \sqrt{\frac{\alpha}{2 |\delta|}} \, \, \mathcal{P}_H (E= \sqrt{\frac{2 |\delta|}{\alpha}}\, , \,n_e)$:
\begin{equation}
\tilde{I} (\Delta_c, n_e) =  \int_{-\infty}^0 I_0(\Delta_c+\delta)  \, \tilde{\mathcal{P}}_H (\delta, n_e) \,  d\delta \quad . 
\label{eq:hm_conv}
\end{equation}
There, $\delta$ is limited to negative values because the Stark shift of Rb $nS_{1/2}$ states is intrinsically negative. The field-free EIT lineshape, $I_0(\Delta_c)$, is measured with the same Ar pressure that is used for the ICP, but with the rf power disabled. In this way, we ensure that collisional shifts and broadening of the EIT line arising from the neutral gas background~\cite{dash2025rydberg} are already accounted for in $I_0(\Delta_c)$. The convolution will then isolate the specific effects of the plasma microfield.

The broadening added by plasma-induced collisions is treated as a homogeneous effect and empirically modeled by convolving $\tilde{I} (\Delta_c, n_e)$ with a Gaussian broadening function, $\mathcal{P}_G(\delta, \gamma) = \exp\left(-\frac{\delta^2}{2\gamma^2}\right) \frac{1} {\sqrt{2 \pi} \gamma}$, in which $\gamma$ is the standard deviation of the collisional broadening. Consequently, the final fit function is given by the three-way convolution:
\begin{equation}
\tilde{\tilde{I}} (\Delta_c, n_e, \gamma) = \mathcal{P}_G (\gamma) \otimes I_0(\Delta_c) \otimes \tilde{\mathcal{P}}_H (n_e) \quad .
\label{eq:all_conv}
\end{equation}

We fit our experimental EIT lineshapes in the plasma using this comprehensive convolution model with the Levenberg-Marquardt algorithm~\cite{newville2016lmfit}. The fit outcomes for the plasma density, $n_e$, and the collisional standard width, $\gamma$, are the main deliverables of the work in characterizing the plasma and the spectral broadening. Procedurally, we first obtain plasma EIT spectra dependent on three main parameters: Ar pressure, $p$, ranging from $5 \text{~mTorr}$ to $50 \text{~mTorr}$, $P_{\text{rf}}$ ranging from $0.25 \text{~W}$ to $5 \text{~W}$, and principal quantum number, $n$, of the Rydberg $nS_{1/2}$ state, varied from $n=25$ to $30$.  The lineshape fits are performed within a window width of $60 \text{~MHz}$ centered about the main EIT peak to exclude distortions from the hyperfine state $F'=3$ of the intermediate state $5P_{3/2}$, which appears approximately $75 \text{~MHz}$ red-detuned from the central EIT resonance in Fig.~\ref{fig:fig2}~(c) at zero $P_{\text{rf}}$. In our present work, the quantitative analysis is restricted to $P_{\text{rf}} < 1.5 \text{~W}$, as the line broadening becomes large at higher powers, which leads to increased fit uncertainties.  The fit results for $n_e$ and $\gamma$, obtained from EIT spectra for the selected case $n=25$, are presented in Fig.~\ref{fig:fig3}.

\begin{figure}
    \centering
    \includegraphics[width=1\linewidth]{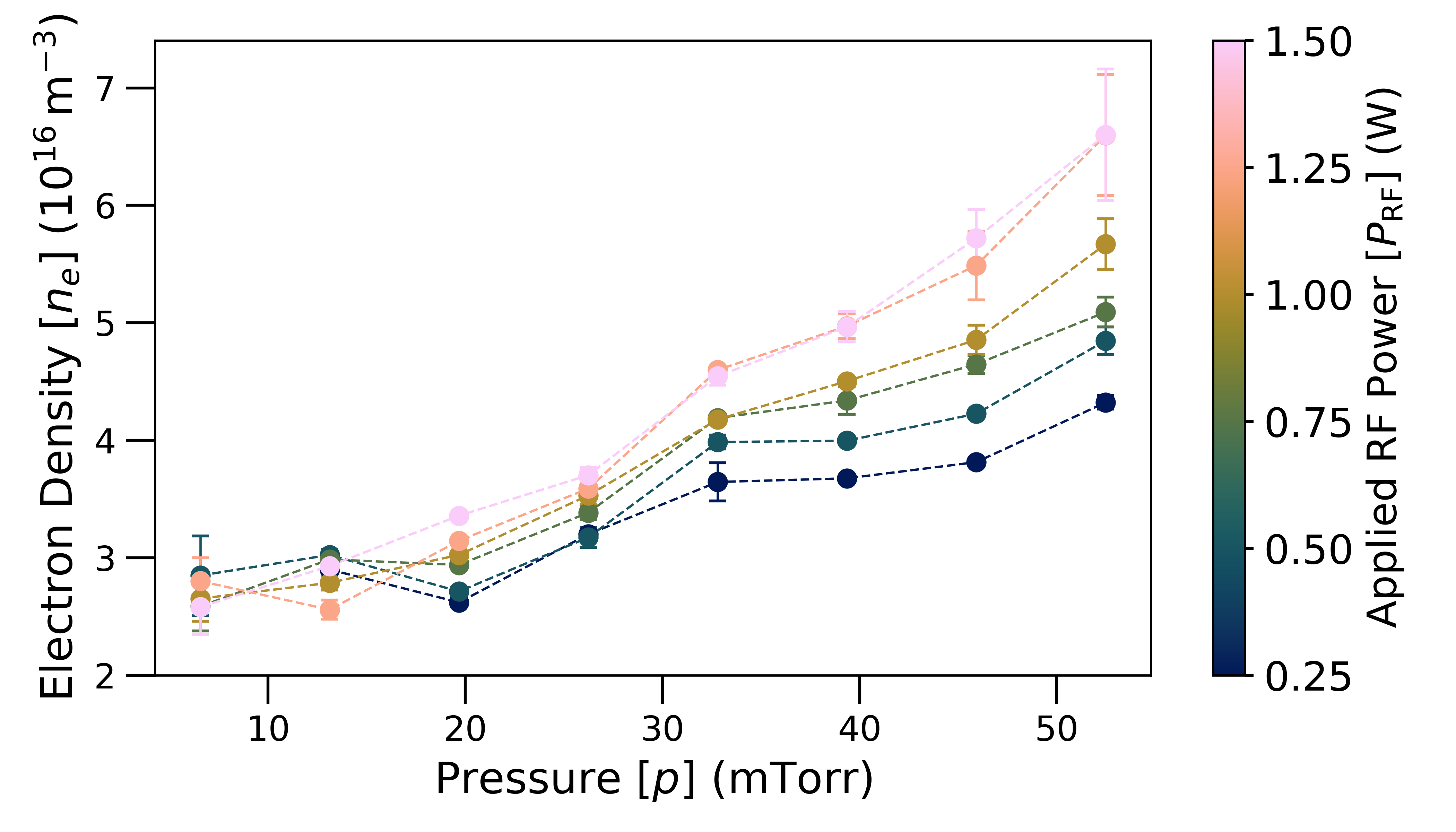}
    \caption{Electron density from Fig.~\ref{fig:fig3}, plotted vs Ar pressure with color-coded powers $P_{\rm{rf}}$. At $p \gtrsim 25$~mTorr, the different-colored curves spread out, reflecting the increase in $\Delta  n_e / \Delta P_{\mathrm{rf}}$. 
}
    \label{fig:fig4}
\end{figure}

As seen in Figs.~\ref{fig:fig3}~(a) and~\ref{fig:fig4}, the rate $\Delta n_e / \Delta P_{\text{rf}}$ at which the plasma density $n_e$ increases with $P_{\text{rf}}$ exhibits a notable uptick at $p \sim 25$~mTorr. This behavior is generally consistent with the energy balance equation for rf plasma, $P_{\text{abs}} \propto n_e$~\cite{lieberman1994principles}. We must caution that $P_{\text{abs}}$, the power absorbed into the plasma, is lower than $P_{\text{rf}}$, the measured power that is delivered into the drive coil~\cite{godyak_2021_Pabs, Godyak_Piejak_Alexandrovich_2002_eedfReview}. The latter, given by the difference between forward and reverse powers reported by the matching network, includes Ohmic losses and may not even be proportional to $P_{\text{abs}}$. The absorbed power $P_{\text{abs}}$ is determined by the intrinsic plasma absorption efficiency, which depends on the plasma parameters. It has been established that power transfer efficiency at $13.56 \text{~MHz}$ is maximal near a critical Ar pressure $p^{*} \approx 25 \text{~mTorr}$, where the electron-neutral collision frequency becomes comparable to the rf drive frequency ~\cite{kralkina2016rf, lieberman1994principles}. Therefore, our observation that the plasma density $n_e$ trends towards a faster and proportional scaling with power for $p \gtrsim p^{*} \approx 25 \text{~mTorr}$ likely results from the ICP entering a regime of improved rf power delivery into the plasma~\cite{yang1999_ICP_Pabs}.

\begin{figure}[h!]
    \centering
    \includegraphics[width=0.95\linewidth]{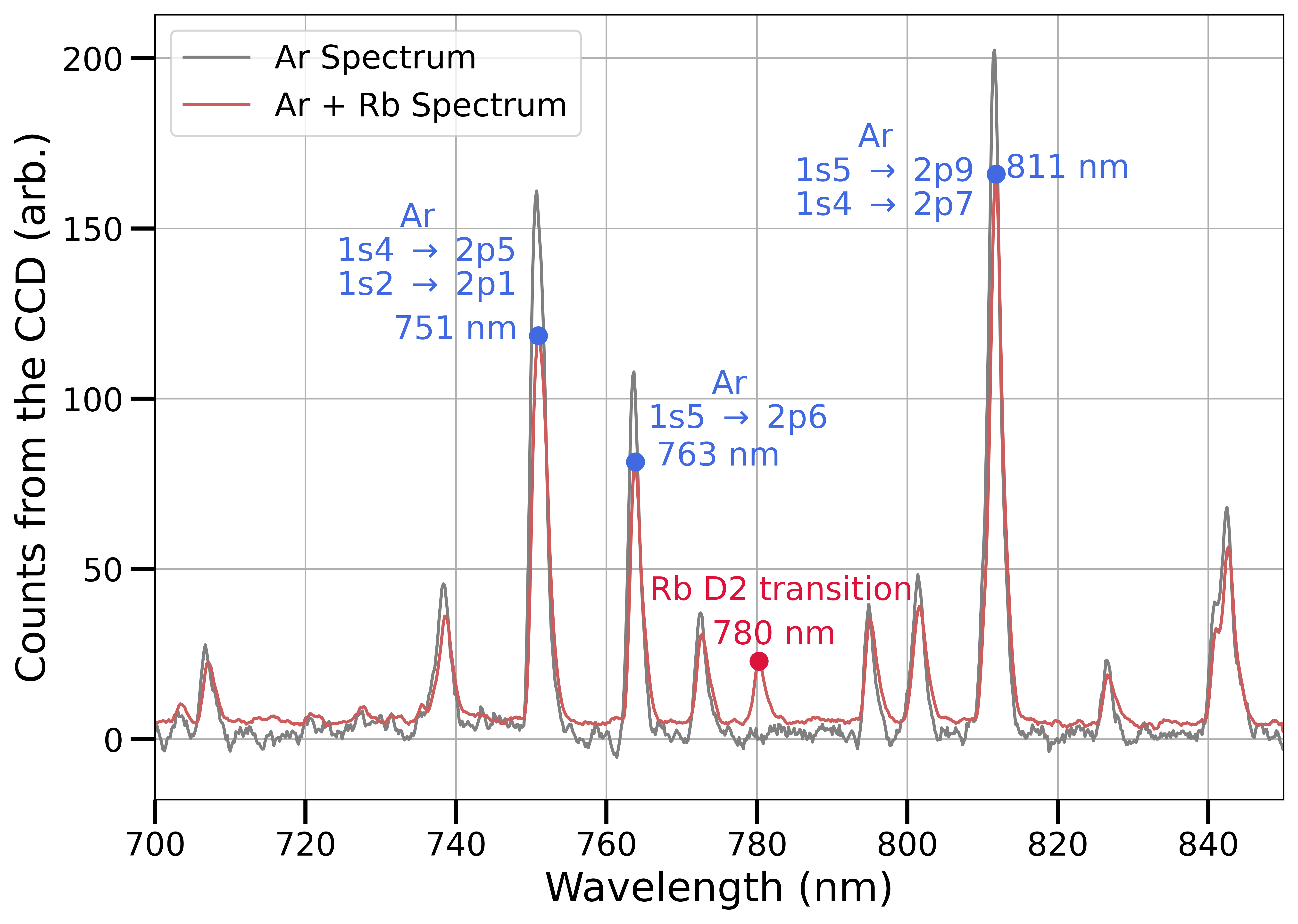}
    \caption{ Optical emission spectrum of the Ar discharge with and without Rb present. The D$_2$ line of Rb at 780~nm is well isolated from the Ar transitions, the strongest of which are labeled (Paschen notation). The Rb emission arises from the fluorescence of Rb vapor in the EIT probe laser. 
    }
    \label{fig:fig5}
\end{figure}

Our plasma density results in Fig.~\ref{fig:fig3}~(a) and~\ref{fig:fig4} are supported by probe-based measurements~\cite{Godyak_Piejak_1998_ne,Zhu_Pu_2007, elfayumi1998hysteresis, Kim_Rao_Cappelli_Sharma_Meyyappan_2001}, which have indicated plasma density values similar to our spectroscopic results under comparable pressure conditions. We note that comparisons across plasma experiments with different probing methods require a scaling of $P_{\text{rf}}$ to plasma-chamber geometries and volumes~\cite{lieberman1994principles}. Further, fit results from EIT spectra with $27S_{1/2}$ and $30S_{1/2}$, not shown, provide similar estimates for plasma density within the experimental uncertainties. The plasma densities from Figs.~\ref{fig:fig3} (a) and \ref{fig:fig4} further yield skin depths $\sim 2~$cm, consistent with our observation in Fig.~\ref{fig:fig2} (d) that the RF field is largely screened from the bulk plasma volume.

We next focus on collisional broadening of EIT lines in the ICP. Fig.~\ref{fig:fig3} (b) shows that our fit results for $\gamma$ exhibit an increasing trend in $P_{\text{rf}}$. The results for $\gamma$ are in the range $\gamma \lesssim 20$~MHz and therefore below the Holtsmark Stark broadening, which ranges $\lesssim 100$~MHz in our work.  As the EIT linewidth in the neutral Ar background gas is already accounted for in the field-free lineshape $I_0(\Delta_c)$ in Eq.~\ref{eq:all_conv}, the broadening $\gamma$ in Fig.~\ref{fig:fig3} (b) arises from plasma collisions, which include Rydberg-atom collisions with electrons, Ar$^+$, and metastable Ar$^*$. Based on previous treatments of $\ell-$mixing~\cite{schiavone1977_electron_lmixing} and ionizing collisions of Rydberg atoms~\cite{beigman1995_collisionReview, griem1967_inelasticbroadening_e_vs_ion}, we assume electron-induced $\ell-$mixing to be dominant due to the $n^5$-scaling of its cross section, followed by inelastic electron scattering, which has cross sections $\propto n^4$ (with principal quantum number $n$). Lesser effects are expected from Ar$^+$ and metastable Ar$^*$ due to lower collision velocities. Quantitatively, our estimates based on~\cite{beigman1995_collisionReview, gallagher} yield collisional linewidths in the range of $15~$MHz for our plasma conditions, which is quite in line with the results in Fig.~\ref{fig:fig3}~(b). Further, in data not shown, fit results for $\gamma$ were seen to rapidly increase with $n$, as expected from the $n^5$ and $n^4$ scalings of Rydberg-atom $\ell$-mixing and inelastic collision cross sections. The predominant role of $\ell$-mixing in collisional Rydberg system has also been noted in ZEKE spectroscopy~\cite{Schlag1998} and cold Rydberg atom experiments~\cite{Dutta2001, Walz2004}.

We finally ascertain that EIT spectroscopy of Rb Rydberg atoms in an Ar ICP is not affected by plasma emission lines from excited Ar species. To prove this, we perform optical emission spectroscopy with an Ocean Optics USB4000 spectrometer in parallel with EIT spectroscopy. The relevant spectral portion of the Ar ICP emission spectrum contains several strong emission lines, as shown in Fig.~\ref{fig:fig5}, but the peak at $\approx$ 780~nm is solely attributed to the fluorescence of Rb atoms from the resonant EIT probe laser. The 780-nm Rb EIT probe peak is well-separated from plasma emission lines and vanishes in the absence of Rb vapor or EIT probe light.

\section{Conclusion}
\label{sec:conclusion}

In summary, we have demonstrated Rydberg-EIT for electric-field diagnostics in low-temperature, low-pressure ICP discharges. The approach leverages the high electric polarizability of Rydberg states, which scales as $n^7$ and is conducive to high electric-field sensitivity. We have measured plasma microfields $\lesssim 10$~V/cm in a continuous ICP, generated in a few mTorr of Ar, over a range of plasma rf drive powers and Ar pressures, and we have observed efficient rf screening from the bulk of the plasma due to the skin effect.
From our work, spectroscopic Rydberg-EIT-based plasma electric-field sensing  emerges as a viable non-invasive method to assess Holtsmark microfields in continuous low-pressure ICP. 
%Rydberg-EIT-based electric-field measurement has considerable potential for spatio-temporally resolved electric-field diagnostics in plasmas. 
%EIT is non-invasive spectroscopic method that allows remote all-optical electric-field measurement, with considerable potential for spatio-temporally resolved electric-field diagnostics in plasmas. 
In future work, electric-field sensitivity may be enhanced by using higher-$n$ Rydberg states. The method may be translated to capacitively coupled rf or dc plasmas, with possible work on spatio-temporal electric-field measurement in plasma sheaths or near dust particles in plasma. In other applications, one may explore the measurement of rf waves in the HF radio band by leveraging the atoms' near-field response to both electric and magnetic fields inside rf load coils.   

\section*{Acknowledgments}
We thank Dr. Alisher Duspayev and Dr. Nithiwadee Thaicharoen for valuable discussions. This project was supported by the U.S. Department of Energy, Office of Science, Office of Fusion Energy Sciences under award number DE-SC0023090. B.D. acknowledges support from the Rackham International Student Fellowship at the University of Michigan. 

\bibliography{refs-fixed}   

\end{document}